\documentclass[prd,preprint,tightenlines,floatfix,showpacs,preprintnumbers,nofootinbib,eqsecnum]{revtex4}

 \usepackage[dvips,final]{graphicx}
  \usepackage{amssymb}
   \usepackage{amsmath}
    \usepackage{amsfonts}
     \usepackage{epsfig}
      \usepackage{bm} 
      
\usepackage{slashed}

\DeclareMathOperator*{\SumInt}{%
\mathchoice%
  {\ooalign{$\displaystyle\sum$\cr\hidewidth$\displaystyle\int$\hidewidth\cr}}
  {\ooalign{\raisebox{.14\height}{\scalebox{.7}{$\textstyle\sum$}}\cr\hidewidth$\textstyle\int$\hidewidth\cr}}
  {\ooalign{\raisebox{.2\height}{\scalebox{.6}{$\scriptstyle\sum$}}\cr$\scriptstyle\int$\cr}}
  {\ooalign{\raisebox{.2\height}{\scalebox{.6}{$\scriptstyle\sum$}}\cr$\scriptstyle\int$\cr}}
}

\begin{document}


\title{Effects of rotation and acceleration in the axial current: density operator vs Wigner function}

\author{George Y. Prokhorov$^1$}
\email{prokhorov@theor.jinr.ru}

\author{Oleg V. Teryaev$^{1,2,3}$}
\email{teryaev@theor.jinr.ru}

\author{Valentin I. Zakharov$^{2,4}$}
\email{vzakharov@itep.ru}
\affiliation{
{$^1$\sl 
Joint Institute for Nuclear Research, Dubna,  Russia \\
$^2$ \sl
Institute of Theoretical and Experimental Physics,
B. Cheremushkinskaya 25, Moscow, Russia \\
$^3$ \sl
National Research Nuclear University MEPhI (Moscow
Engineering Physics Institute), Kashirskoe Shosse 31, 115409 Moscow,
Russia
 \\
$^4$ \sl
School of Biomedicine, Far Eastern Federal University, 690950 Vladivostok, Russia
}\vspace{1.5cm}
}

\begin{abstract}
\vspace{0.5cm}
The hydrodynamic coefficients in the axial current are calculated on the basis of the equilibrium quantum statistical density operator in the third order of perturbation theory in thermal vorticity tensor both for the case of massive and massless fermions. The coefficients obtained describe third-order corrections to the Chiral Vortical Effect and include the contribution from local acceleration. We show that the methods of the Wigner function and the statistical density operator lead to the same result for an axial current in describing effects associated only with vorticity when the local acceleration is zero, but differ in describing mixed effects for which both acceleration and vorticity are significant simultaneously.
\end{abstract}


\maketitle

\section{Introduction}

Increasing attention to relativistic hydrodynamics from the experimental point of view is due to the fact that after the collision of heavy ions a quark-gluon plasma cluster is formed.
At the theoretical level, different remarkable effects associated with the properties of relativistic fluids are discovered. The two most famous effects of this kind are the Chiral Magnetic (CME) \cite{Son:2009tf, Kharzeev:2012ph, Fukushima:2008xe, Sadofyev:2010is, 
Zakharov:2012vv} and the Chiral Vortical Effect (CVE) \cite{Vilenkin:1979ui, Vilenkin:1980zv, Son:2009tf, Kharzeev:2012ph, Sadofyev:2010is, 
Zakharov:2012vv, Buzzegoli:2017cqy, Gao:2017gfq, Landsteiner:2012kd, Prokhorov:2017atp, Prokhorov:2018qhq, Zakharov:2016lhp}, which will be discussed below. The appearance of baryon polarization in collisions of heavy ions can be one of the important experimental consequences of CVE, as was shown in \cite{Rogachevsky:2010ys, Sorin:2016smp, Baznat:2017jfj, Baznat:2017ars} and \cite{Becattini:2017gcx, Becattini:2016gvu, Karpenko:2017lyj}.

Various theoretical methods for investigation of the chiral effects associated with the nonuniform motion of the medium have been developed: within the framework of field theory at finite temperatures in rotating systems \cite{Vilenkin:1980zv}, in the framework of hydrodynamics with the axial anomaly \cite{Son:2009tf}, from an axial anomaly in effective field theory \cite{Sadofyev:2010is, 
Zakharov:2012vv}, etc. All of these approaches show the existence of CVE, which is thus a well-theoretically grounded effect.

However, the issue of higher-order corrections with respect to derivatives to this effect remains open. If the first-order term with respect to the angular velocity is related to the axial electromagnetic anomaly \cite{Son:2009tf, Sadofyev:2010is, Zakharov:2012vv}, then higher-order terms should be related to other anomalies in quantum field theory, in particular, to the gravitational anomaly. Thus, the study of corrections of higher orders will make it possible will improve our understanding of the effect of the anomalies of quantum field theory on relativistic hydrodynamics.

Another open question relates to the study of effects associated with acceleration in chiral phenomena. In particular, these effects were discussed in \cite{Zakharov:2016lhp, Prokhorov:2017atp, Prokhorov:2018qhq, Buzzegoli:2017cqy, Becattini:2015nva, Florkowski:2018myy, Becattini:2017ljh, Stone:2018zel}. In \cite{Zakharov:2016lhp} it is shown that their occurrence is dictated by the principle of equivalence. In \cite{Prokhorov:2018qhq, Florkowski:2018myy, Becattini:2017ljh} the relationship of these effects to the Unruh effect is found. In particular, it is shown in \cite{Prokhorov:2018qhq, Florkowski:2018myy, Becattini:2017ljh} that Unruh temperature appears as a boundary temperature for chiral effects.

In this paper we will touch on both of these issues. We will be interested in two recently developed methods for investigation of chiral effects: the first of them is based on the ansatz of the Wigner function \cite{Becattini:2013fla, Florkowski:2018myy, Florkowski:2018ahw, Prokhorov:2017atp, Prokhorov:2018qhq} (recently in \cite{Florkowski:2018ahw} it was shown that this Wigner function satisfies the zeroth-order kinetic equation with the vanishing collision term), the second approach is based on the equilibrium quantum statistical density operator \cite{Zubarev, Weert, Becattini:2014yxa, Hayata:2015lga, Buzzegoli:2017cqy,  Becattini:2015nva, Becattini:2012tc, Becattini:2017ljh, Hongo:2016mqm}. The purpose of this paper is to compare these two approaches in describing higher order effects at the equilibrium mean value of the axial current.

In \cite{Becattini:2013fla} an ansatz of the Wigner function was proposed, taking into account the effects associated with thermal vorticity. In \cite{Buzzegoli:2017cqy, Prokhorov:2017atp, Prokhorov:2018qhq} an axial current was calculated on the basis of this Wigner function, and the resulting expression for the current exactly coincides with the standard formula for CVE. In \cite{Prokhorov:2018qhq} it was shown that in the expression for the mean value of the axial current, the angular velocity and acceleration play the role of additional chemical potentials, and the acceleration corresponds to an imaginary chemical potential. In particular, with parallel vorticity and acceleration, a combination of the form $\mu \pm (\Omega \pm ia)/2$ (where $\Omega$ and $a$ are the modules of the three-dimensional angular velocity and acceleration, respectively, in the comoving frame of reference) appeared in Fermi distribution. Indications that the angular velocity plays the role of an additional chemical potential were also obtained in \cite{Florkowski:2017ruc}.

In \cite{Prokhorov:2017atp, Prokhorov:2018qhq}, corrections of higher orders to CVE were investigated, and it was shown that the axial current contains a third-order term with respect to the angular velocity. The corresponding term with the same coefficient appeared in \cite{Vilenkin:1979ui, Vilenkin:1980zv}. It is also shown that the current contains third-order terms with respect to derivatives, quadratic in terms of local acceleration.

In this paper, we will use an independent approach based on the quantum statistical density operator for a medium with thermal vorticity \cite{Zubarev, Weert, Becattini:2014yxa, Hayata:2015lga, Buzzegoli:2017cqy, Becattini:2015nva, Becattini:2012tc, Becattini:2017ljh, Hongo:2016mqm} using the calculation technique developed in \cite{Buzzegoli:2017cqy,Becattini:2015nva}. In \cite{Zubarev, Weert, Becattini:2014yxa, Hayata:2015lga, Buzzegoli:2017cqy, Becattini:2015nva, Becattini:2012tc, Becattini:2017ljh, Hongo:2016mqm} it is shown, that a moving medium is described by a density operator containing an additional term, in comparison with a grand canonical distribution. See also recent paper \cite{Buzzegoli:2018wpy}, where the effects associated with the axial chemical potential were investigated. In \cite{Buzzegoli:2017cqy} the mean value of the axial current in the linear approximation in the thermal vorticity for free Dirac fields was calculated and it was shown that it coincides with the prediction resulting from the Wigner function \cite{Becattini:2013fla, Prokhorov:2017atp, Prokhorov:2018qhq}, that is, both these methods lead to a standard formula for CVE. We will see later that these two methods coincide in describing the effects associated with rotation separately, but differ in describing the mixed effects associated with acceleration and rotation \footnote{We are grateful to E. Grossi who pointed out this fact.}.

We calculate the hydrodynamic coefficients in the third order of perturbation theory following \cite{Buzzegoli:2017cqy,Becattini:2015nva} for free Dirac fields and compare the resulting expression with the result of the approach based on the Wigner function. The two methods agree with each other when considering the rotation of the system without acceleration in the comoving reference system in the general case of massive fermions in the third order in thermal vorticity and differ when considering mixed effects associated with acceleration and rotation simultaneously.

The system of units $\hbar=c=k=1$ is used.

\section{Analysis of the effects of non-uniform motion of the medium in the axial current on the basis of the equilibrium density operator}
\label{Sec:main}

Following \cite{Zubarev, Weert, Becattini:2014yxa, Hayata:2015lga, Buzzegoli:2017cqy, Becattini:2015nva, Becattini:2012tc, Becattini:2017ljh, Hongo:2016mqm} a medium in the state of local thermodynamic equilibrium is described by the covariant quantum density operator of the next form
\begin{eqnarray}
&&\hat{\rho}=\frac{1}{Z}\exp\Big\{
-\int_{\Sigma}d\Sigma_{\mu}[\hat{T}^{\mu\nu}(x)\beta_{\nu}(x)-\zeta(x) \hat{j}^{\mu}(x)
]\Big\} \,,
\label{rho}
\end{eqnarray}
where the integration over the 3-dimensional hypersurface $\Sigma$ is performed. Here $\beta_{\mu}=\frac{u_{\mu}}{T}$ is the 4-vector of the inverse temperature, $T$ is the temperature in the comoving frame, $\zeta=\frac{u}{T}$ is the ratio of the chemical potential in the comoving reference system to the temperature, $\hat{T}^{\mu\nu}$ and $\hat{j}^{\mu}$ are the energy-momentum tensor and current operators. The general conditions of the global thermodynamic equilibrium for a medium with rotation and acceleration, under which the density operator (\ref{rho}) ceases to depend on the choice of the hypersurface $\Sigma$, over which the integration takes place, have the form \cite{Becattini:2012tc, DeGroot:1980dk, Buzzegoli:2017cqy, Becattini:2015nva, Florkowski:2018ahw}
\begin{eqnarray}
\beta_{\mu}= b_{\mu}+\varpi_{\mu\nu}x_{\nu}\,,\quad b_{\mu}=\mathrm{const}\,,\quad\, \varpi_{\mu\nu}=\mathrm{const}\,,\quad\varpi_{\mu\nu}= -\frac{1}{2}(\partial_{\mu}\beta_{\nu}-\partial_{\nu}\beta_{\mu})\,,
\label{global}
\end{eqnarray}
where $\varpi_{\mu\nu}$ is the thermal vorticity tensor. The thermal vorticity tensor $\varpi_{\mu\nu}$ contains information about local acceleration and rotation in the system, which corresponds to its expansion into the thermal acceleration vector $\alpha_{\mu}$ and the pseudovector of thermal vorticity $w_{\mu}$
\begin{eqnarray}
\varpi_{\mu\nu}= \epsilon_{\mu\nu\alpha\beta}w^{\alpha}u^{\beta}+\alpha_{\mu}u_{\nu}-\alpha_{\nu}u_{\mu}\,.
\label{wdec}
\end{eqnarray}

In the state of global equilibrium (\ref{global}) the thermal acceleration and vorticity become proportional to the corresponding kinematic acceleration $a_{\mu}$ and vorticity $\omega_{\mu}$
\begin{eqnarray}
w_{\mu}=\frac{\omega_{\mu}}{T}=
\frac{1}{2T}
\epsilon_{\mu\nu\alpha\beta}u^{\nu}\partial^{\alpha}u^{\beta}\,,\quad
\alpha_{\mu}=\frac{a_{\mu}}{T}=\frac{1}{T}u^{\nu}\partial_{\nu}u_{\mu}\,.
\label{wa}
\end{eqnarray}

Under the condition (\ref{global}) the density operator (\ref{rho}) takes the form of an equilibrium density operator \cite{Buzzegoli:2017cqy, Becattini:2015nva, Becattini:2017ljh}
\begin{eqnarray}
&&\hat{\rho}=\frac{1}{Z}\exp\Big\{-\beta_{\mu}(x)\hat{P}^{\mu}
+\frac{1}{2}\varpi_{\mu\nu}\hat{J}^{\mu\nu}_x+\zeta \hat{Q}
\Big\} \,,
\label{rho global}
\end{eqnarray}
where $\hat{P}$ is the 4-momentum operator, $\hat{Q}$ is the charge operator, and $\hat{J}_x$ are the generators of the Lorentz transformations displaced to the point $x$
\begin{eqnarray}
&& \hat{J}^{\mu\nu}_x=
\int d\Sigma_{\lambda}\big[(y^{\mu}-x^{\mu})\hat{T}^{\lambda\nu}(y)-(y^{\nu}-x^{\nu})\hat{T}^{\lambda\mu}(y)\big] \,.
\label{J}
\end{eqnarray}

The technique for calculating the mean values of physical quantities on the basis of (\ref{rho global}) was developed in the papers \cite{Buzzegoli:2017cqy, Becattini:2015nva}, in which hydrodynamic coefficients were calculated in the second order in the thermal vorticity tensor in various observables for scalar and Dirac fields. We will follow the calculation algorithm proposed in \cite{Buzzegoli:2017cqy, Becattini:2015nva} and obtain third-order corrections in the thermal vorticity tensor. Note that according to \cite{Vilenkin:1980zv, Vilenkin:1979ui, Prokhorov:2017atp, Prokhorov:2018qhq} in the massless limit, all the terms in the axial current above third-order in the thermal vorticity tensor are canceled (at least at a temperature above Unruh temperature); therefore, there are reasons to believe that corrections above the third order will be zero for the massless case also in this method.

The mean value of an operator of a physical quantity can be calculated using (\ref{rho global}) according to formula
\begin{eqnarray}
&&\langle\hat{O}(x)\rangle = \mathrm{tr}\{\hat{\rho}\hat{O}(x)\}_{\mathrm{ren}} \,,
\label{mean}
\end{eqnarray}
where $ren$ denotes the renormalization procedure. Following \cite{Buzzegoli:2017cqy} and expanding (\ref{rho global}) into a series of thermal vorticity, we obtain the following expression for the axial current in the third order of perturbation theory
\begin{eqnarray}  \label{ja3}
&& \langle\hat{j}_5^{\lambda}(x)\rangle =
\frac{\varpi_{\mu\nu}}{2|\beta|}\int^{|\beta|}_{0} d\tau\langle T_{\tau}\,\hat{J}^{\mu\nu}_{-i\tau u}\hat{j}_5^{\mu}(0)\rangle_{\beta(x),c}+ \\ \nonumber
&&\frac{\varpi_{\mu\nu}\varpi_{\rho\sigma}\varpi_{\alpha\beta}}{48|\beta|^3}\int^{|\beta|}_{0} d\tau_1d\tau_2d\tau_3\langle T_{\tau}\,\hat{J}^{\mu\nu}_{-i\tau_1 u}\hat{J}^{\rho\sigma}_{-i\tau_2 u}\hat{J}^{\alpha\beta}_{-i\tau_3 u}\hat{j}_5^{\lambda}(0)\rangle_{\beta(x),c} + O(\varpi^5)\,,
\end{eqnarray}
where all operators must be expressed through Dirac fields using standard formulas. In (\ref{ja3}) only connected correlators enter, since all disconnected correlators are canceled due to the contribution of the denominator $1/Z$ in (\ref{rho global}). This fact is shown in the lower index $c$, the lower index $\beta(x)$ means that the mean values are taken at $\varpi=0$, that is, the averaging is performed over the grand canonical distribution. $T_{\tau}$ means the ordering of operators with respect to the imaginary time $\tau$, and $|\beta|=\frac{1}{T}$. The contributions of the zero and the second order in (\ref{ja3})  are zero, which is connected with the requirement of parity equality in both parts of the equation, therefore in the third order of perturbation theory
\begin{eqnarray}
&& \langle\hat{j}_5^{\lambda}(x)\rangle =\langle\hat{j}_5^{\lambda}(x)\rangle^{(1)}+\langle\hat{j}_5^{\lambda}(x)\rangle^{(3)}+O(\varpi^5)\,.
\label{jafull}
\end{eqnarray}

The first-order contribution to (\ref{jafull}) was calculated in  \cite{Buzzegoli:2017cqy} Eq. (7.4)
\begin{eqnarray} 
&& \langle\hat{j}_5^{\lambda}(x)\rangle^{(1)} =
-\frac{1}{2\pi^2}\int^{\infty}_{0}dp\,p^2 \Big(n'_F(E_p-\mu)+n'_F(E_p+\mu)\Big)\omega^{\lambda}\,,
\label{ja1}
\end{eqnarray}
where the energy derivative is taken $\frac{d}{dE_p}$ and $E_p=\sqrt{p^2+m^2}$ as usual and $p^2=\bold{p}^2$. Let us calculate the third-order corrections in (\ref{jafull}).  Parity allows the appearance of terms of three types
\begin{eqnarray} 
&& \langle\hat{j}_5^{\lambda}(x)\rangle^{(3)} =
A_1w^2w^{\lambda}+A_2\alpha^2 w^{\lambda}+A_3(w\alpha)\alpha^{\lambda}\,.
\label{Afirst}
\end{eqnarray}

We note that in the presence of an axial chemical potential, additional tensor structures appear, not included in (\ref{Afirst}), according to \cite{Buzzegoli:2018wpy}. In what follows it is convenient to introduce the operators of boost $\hat{K}$ and angular momentum $\hat{J}$
\begin{eqnarray}
&& \hat{J}^{\mu\nu}=u^{\mu}\hat{K}^{\nu}-u^{\nu}\hat{K}^{\mu}-
\epsilon^{\mu\nu\rho\sigma}u_{\rho}\hat{J}_{\sigma} \,.
\label{Jdec}
\end{eqnarray}

Substituting (\ref{Jdec}) and (\ref{wdec}) into (\ref{ja3}),and again using the parity arguments, we get
\begin{eqnarray}
&&\langle\hat{j}_5^{\lambda}(x)\rangle^{(3)}=-\frac{1}{6|\beta|^3}\Big(
\alpha_{\mu}w_{\nu}\alpha_{\rho}
\int^{|\beta|}_0 d\tau_1 d\tau_2 d\tau_3\langle T_{\tau} 
\big\{\hat{K}^{\mu}_{-i\tau_1 u},\hat{J}^{\nu}_{-i\tau_2 u}\big\}\hat{K}^{\rho}_{-i\tau_3 u}
\hat{j}_5^{\lambda}(0)\rangle_{\beta(x),c}
+ \nonumber \\
&&\alpha_{\mu}\alpha_{\nu}w_{\rho}
\int^{|\beta|}_0 d\tau_1 d\tau_2 d\tau_3\langle T_{\tau} 
\hat{K}^{\mu}_{-i\tau_1 u}\hat{K}^{\nu}_{-i\tau_2 u}\hat{J}^{\rho}_{-i\tau_3 u}
\hat{j}_5^{\lambda}(0)\rangle_{\beta(x),c}
+\nonumber \\
&& w_{\mu}w_{\nu}w_{\rho}
\int^{|\beta|}_0 d\tau_1 d\tau_2 d\tau_3\langle T_{\tau} 
\hat{J}^{\mu}_{-i\tau_1 u}\hat{J}^{\nu}_{-i\tau_2 u}\hat{J}^{\rho}_{-i\tau_3 u}
\hat{j}_5^{\lambda}(0)\rangle_{\beta(x),c}
 \Big)\,.
\label{ja3KJ}
\end{eqnarray}

Comparing (\ref{ja3KJ}) with (\ref{Afirst}), and taking into account the independence of the coefficients $A_1$, $A_2$, $A_3$ from the choice of the frame of reference and the specific form $\varpi$  (or using the expansion for correlators \cite{Buzzegoli:2017cqy}), it is possible to express the coefficients in terms of quantum correlators
\begin{eqnarray}
&&A_1=-\frac{1}{6|\beta|^3}
\int^{|\beta|}_0 d\tau_1 d\tau_2 d\tau_3
\langle T_{\tau} 
\hat{J}^{3}_{-i\tau_1 u}\hat{J}^{3}_{-i\tau_2 u}\hat{J}^{3}_{-i\tau_3 u}
\hat{j}_5^{3}(0)\rangle_{\beta(x),c} \,,\nonumber \\
&&A_2=-\frac{1}{6|\beta|^3}\Big(
\int^{|\beta|}_0 d\tau_1 d\tau_2 d\tau_3
\langle T_{\tau} 
\big(\hat{K}^{1}_{-i\tau_1 u}\hat{J}^{3}_{-i\tau_2 u}+\hat{J}^{3}_{-i\tau_1 u}\hat{K}^{1}_{-i\tau_2 u}\big)
\hat{K}^{1}_{-i\tau_3 u}
\hat{j}_5^{3}(0)\rangle_{\beta(x),c}
+ \nonumber \\
&&\int^{|\beta|}_0 d\tau_1 d\tau_2 d\tau_3
\langle T_{\tau} 
\hat{K}^{1}_{-i\tau_1 u}\hat{K}^{1}_{-i\tau_2 u}\hat{J}^{3}_{-i\tau_3 u}
\hat{j}_5^{3}(0)\rangle_{\beta(x),c}\Big)
 \,,\nonumber \\
&&A_3=-\frac{1}{6|\beta|^3}\Big(
\int^{|\beta|}_0 d\tau_1 d\tau_2 d\tau_3
\langle T_{\tau} 
\big(\hat{K}^{3}_{-i\tau_1 u}\hat{J}^{3}_{-i\tau_2 u}+\hat{J}^{3}_{-i\tau_1 u}\hat{K}^{3}_{-i\tau_2 u}\big)\hat{K}^{3}_{-i\tau_3 u}
\hat{j}_5^{3}(0)\rangle_{\beta(x),c}
+ \nonumber \\
&&\int^{|\beta|}_0 d\tau_1 d\tau_2 d\tau_3
\langle T_{\tau} 
\hat{K}^{3}_{-i\tau_1 u}\hat{K}^{3}_{-i\tau_2 u}\hat{J}^{3}_{-i\tau_3 u}
\hat{j}_5^{3}(0)\rangle_{\beta(x),c}\Big)-A_2\,.
\label{Asecond}
\end{eqnarray}

Expressing the operators $\hat{K}$ and $\hat{J}$ in terms of the energy-momentum tensor $\hat{T}^{\mu\nu}$ using the formulas (\ref{Jdec}), (\ref{J}), we reduce the calculation of the coefficients in (\ref{Asecond}) to the calculation of correlators of the form
\begin{eqnarray}
&& C^{\alpha_1\alpha_2|\alpha_3\alpha_4|\alpha_5\alpha_6|\lambda|ijk}=
\frac{1}{|\beta|^3}\int d\tau_x d\tau_y d\tau_z d^3x d^3y d^3z
\langle
T_{\tau}\hat{T}^{\alpha_1\alpha_2}(X) \nonumber \\
&&\hat{T}^{\alpha_3\alpha_4}(Y)\hat{T}^{\alpha_5\alpha_6}(Z)j_5^{\lambda}(0)\rangle_{\beta(x),c}x^i y^j z^k \,,
\label{C}
\end{eqnarray}
here $X=(\tau_x,\bold{x})$. The corresponding expressions for the coefficients
\begin{eqnarray} \nonumber
&& A_1=-\frac{1}{6}\Big\{
C^{02|02|02|3|111}+C^{02|01|01|3|122}+C^{01|02|01|3|212}+C^{01|01|02|3|221} \\ \nonumber
&&-C^{01|01|01|3|222}-C^{01|02|02|3|211}-C^{02|01|02|3|121}-C^{02|02|01|3|112}\Big\} \,, \\ \nonumber
&& A_2=-\frac{1}{6}\Big\{
C^{02|00|00|3|111}+C^{00|02|00|3|111}+C^{00|00|02|3|111}-C^{01|00|00|3|211} \\ \nonumber
&&-C^{00|01|00|3|121}-C^{00|00|01|3|112}\Big\} \,, \\ \nonumber
&& A_3=-A_2-\frac{1}{6}\Big\{
C^{02|00|00|3|133}+C^{00|02|00|3|313}+C^{00|00|02|3|331} \\
&&-C^{01|00|00|3|233}-C^{00|01|00|3|323}-C^{00|00|01|3|332}\Big\} \,.
\label{AC}
\end{eqnarray}

Thus, the calculation of the coefficients in (\ref{Afirst}) reduces to calculation of correlators of the form (\ref{C}). Correlators (\ref{C}) can be calculated by analogy with the way it was done in \cite{Buzzegoli:2017cqy} in calculating first-order and second-order hydrodynamic coefficients. The derivation of the formulas (\ref{Csol}), (\ref{Bsol}) is given in Appendix \ref{Sec:correlators}
\begin{eqnarray} \nonumber
&& C^{\alpha_1\alpha_2|\alpha_3\alpha_4|\alpha_5\alpha_6|\lambda|ijk}=
-\frac{i}{128\pi^3|\beta|^3}\int\sum_{\def\arraystretch{0.5}\begin{array}{ll}
{\scriptscriptstyle s_1,s_2,s_3,}\vspace{0.1mm}\\
{\scriptscriptstyle s_4=\pm 1}
\end{array}} d\tau_x d\tau_y d\tau_z p^2 dp \sin(\theta) d\theta d\varphi \\ \nonumber
&&\Big[\Big(\frac{\partial^3}{\partial r^{k}\partial k^{j}\partial p^{i}}+\frac{\partial^3}{\partial r^{k}\partial k^{j}\partial k^{i}}\Big)
B_{-+-+,(\tau_x-\tau_y),(\tau_x-\tau_z)}^{\alpha_3\alpha_4\alpha_1\alpha_2\alpha_5\alpha_6\lambda}(\widetilde{K},\widetilde{K},-\widetilde{P},\widetilde{P},\widetilde{P},\widetilde{Q},-\widetilde{Q},-\widetilde{Q},\widetilde{R},-\widetilde{R})+ \\ \nonumber
&&\Big(\frac{\partial^3}{\partial r^{k}\partial k^{i}\partial p^{j}}+\frac{\partial^3}{\partial r^{k}\partial k^{i}\partial k^{j}}\Big)
B_{+-+-,(\tau_x-\tau_y),(\tau_x-\tau_z)}^{\alpha_5\alpha_6\alpha_3\alpha_4\alpha_1\alpha_2\lambda}(\widetilde{R},\widetilde{R},-\widetilde{Q},\widetilde{Q},\widetilde{Q},\widetilde{P},-\widetilde{P},-\widetilde{P},\widetilde{K},-\widetilde{K})+ \\ \nonumber
&&\Big(\frac{\partial^3}{\partial r^{k}\partial k^{j}\partial p^{i}}+\frac{\partial^3}{\partial r^{k}\partial k^{j}\partial k^{i}}\Big)
B_{+-+-,(\tau_x-\tau_y),(\tau_x-\tau_z)}^{\alpha_5\alpha_6\alpha_1\alpha_2\alpha_3\alpha_4\lambda}(\widetilde{R},\widetilde{R},-\widetilde{Q},\widetilde{Q},\widetilde{Q},\widetilde{P},-\widetilde{P},-\widetilde{P},\widetilde{K},-\widetilde{K})+ \\ \nonumber
&&\Big(\frac{\partial^3}{\partial r^{i}\partial k^{j}\partial p^{k}}+\frac{\partial^3}{\partial r^{i}\partial k^{j}\partial k^{k}}\Big)
B_{-+-+,(\tau_x-\tau_y),(\tau_x-\tau_z)}^{\alpha_3\alpha_4\alpha_5\alpha_6\alpha_1\alpha_2\lambda}(\widetilde{K},\widetilde{K},-\widetilde{P},\widetilde{P},\widetilde{P},\widetilde{Q},-\widetilde{Q},-\widetilde{Q},\widetilde{R},-\widetilde{R})+ \\ \nonumber
&&\Big(\frac{\partial^3}{\partial r^{k}\partial k^{i}\partial p^{j}}+\frac{\partial^3}{\partial r^{k}\partial k^{i}\partial k^{j}}\Big)
B_{-+-+,(\tau_x-\tau_y),(\tau_x-\tau_z)}^{\alpha_1\alpha_2\alpha_3\alpha_4\alpha_5\alpha_6\lambda}(\widetilde{K},\widetilde{K},-\widetilde{P},\widetilde{P},\widetilde{P},\widetilde{Q},-\widetilde{Q},-\widetilde{Q},\widetilde{R},-\widetilde{R})+ \\ \nonumber
&&\Big(\frac{\partial^3}{\partial r^{j}\partial k^{i}\partial p^{k}}+\frac{\partial^3}{\partial r^{j}\partial k^{i}\partial k^{k}}\Big)
B_{-+-+,(\tau_x-\tau_y),(\tau_x-\tau_z)}^{\alpha_1\alpha_2\alpha_5\alpha_6\alpha_3\alpha_4\lambda}(\widetilde{K},\widetilde{K},-\widetilde{P},\widetilde{P},\widetilde{P},\widetilde{Q},-\widetilde{Q},-\widetilde{Q},\widetilde{R},-\widetilde{R})
\Big]\\
&&\frac{1}{E_{p}E_{q}E_{k}E_{r}}
e^{(\tau_x-\tau_y)s_1 E_{p}+(\tau_x-\tau_z)s_2 E_{q}+\tau_y s_3 E_{k}+\tau_z s_4 E_{r}} \Big|\def\arraystretch{0.5}
\begin{array}{ll}
{\scriptstyle \bold{q}=-\bold{p}}\\
{\scriptstyle \bold{k}=\bold{p}}\\
{\scriptstyle \bold{r}=-\bold{p}}
\end{array}\,.
\label{Csol}
\end{eqnarray}

Here, following \cite{Buzzegoli:2017cqy}, we introduce the notation  $\widetilde{P}=\widetilde{P}(s_1)=(-is_1 E_p,\bold{p})$, and accordingly we have $\widetilde{Q}=\widetilde{Q}(s_2)$, $\widetilde{K}=\widetilde{K}(s_3)$, $\widetilde{R}=\widetilde{R}(s_4)$. The derivatives act on the whole expression to the right of them. The quantities $B_{g_1 g_2 g_3 g_4,\tau_1,\tau_2}^{\alpha_1\alpha_2\alpha_3\alpha_4\alpha_5\alpha_6\lambda}(\{P\})$, are defined by the formula
\begin{eqnarray} \nonumber
&&B_{g_1 g_2 g_3 g_4,\tau_1,\tau_2}^{\alpha_1\alpha_2\alpha_3\alpha_4\alpha_5\alpha_6\lambda}(\{P\})= \frac{1}{64}i^{\delta_{0\alpha_1}+\delta_{0\alpha_2}+\delta_{0\alpha_3}+
\delta_{0\alpha_4}+\delta_{0\alpha_5}+\delta_{0\alpha_6}+\delta_{0\lambda}-1} S_{\alpha_1\alpha_2}S_{\alpha_3\alpha_4}S_{\alpha_5\alpha_6}\\ \nonumber
&&(iP_3^{\alpha_2}-iP_2^{\alpha_2})
(iP_6^{\alpha_4}-iP_5^{\alpha_4})
(iP_9^{\alpha_6}-iP_8^{\alpha_6})
\Big[i P_1^{\alpha_7} i P_4^{\alpha_8}i P_7^{\alpha_9}i P_{10}^{\alpha_{10}}\mathrm{tr}_5^{7,1,8,3,9,5,10,\lambda}+\\ \nonumber
&&m^2iP_1^{\alpha_7}iP_4^{\alpha_8}\mathrm{tr}_5^{7,1,8,3,5,\lambda}+
m^2iP_1^{\alpha_7}iP_7^{\alpha_9}\mathrm{tr}_5^{7,1,3,9,5,\lambda}+
m^2iP_1^{\alpha_7}iP_{10}^{\alpha_{10}}\mathrm{tr}_5^{7,1,3,5,10,\lambda}+\\ \nonumber
&&m^2iP_4^{\alpha_8}iP_7^{\alpha_9}\mathrm{tr}_5^{1,8,3,9,5,\lambda}+
m^2iP_4^{\alpha_8}iP_{10}^{\alpha_{10}}\mathrm{tr}_5^{1,8,3,5,10,\lambda}+
m^2iP_7^{\alpha_9}iP_{10}^{\alpha_{10}}\mathrm{tr}_5^{1,3,9,5,10,\lambda}+\\ \nonumber
&&m^4 \mathrm{tr}_5^{1,3,5,\lambda}\Big] 
\big\{\theta(-s_1\tau_1)-n_F(E_p+s_1 g_1\mu)\big\}\big\{\theta(-s_2\tau_2)-n_F(E_q+s_2 g_2\mu)\big\}\\ 
&&\big\{\theta(-s_3)-n_F(E_k+s_3 g_3\mu)\big\}
\big\{\theta(-s_4)-n_F(E_r+s_4 g_4\mu)\big\}\,. 
\label{Bsol}
\end{eqnarray}

Here we introduce the operator $S_{\alpha\beta}$, which symmetrizes the expression following it, so that $S_{\alpha\beta} f_{\alpha\beta}= f_{\alpha\beta}+f_{\beta\alpha}$. The trace of an arbitrary number of Euclidean Dirac matrices $\tilde{\gamma}_{\mu}= i^{1-\delta_{0\mu}}\gamma_{\mu}$ \cite{Laine:2016hma} we denoted by
$\mathrm{tr}(\tilde{\gamma}^{\alpha_{n_1}}\tilde{\gamma}^{\alpha_{n_2}}...\tilde{\gamma}^{\alpha_{n_N}}\tilde{\gamma}^{\lambda}\tilde{\gamma}^{5})
=\mathrm{tr}_5^{n_1,n_2,...,n_N,\lambda}$.

Using the formulas (\ref{Csol}), (\ref{Bsol}), we can now calculate the coefficients $A_1$, $A_2$, $A_3$, performing the remaining operations of integration and differentiation explicitly. Omitting the intermediate calculations, we give the final result in the general case $m\neq 0$
\begin{eqnarray} \nonumber
&& A_1=\frac{1}{48\pi^2|\beta|^3}
\int_{0}^{\infty}dp\Big(n'''_F(E_p-\mu)+n'''_F(E_p+\mu)\Big)p^2
\,, \\ \nonumber
&& A_2=\frac{1}{16\pi^2|\beta|^3}
\int_{0}^{\infty}dp\Big(n'''_F(E_p-\mu)+n'''_F(E_p+\mu)\Big)(p^2+\frac{m^2}{3})
\,, \\
&& A_3=0\,,
\label{Asol}
\end{eqnarray}
where the derivative of the third order in energy is taken $\frac{d^3}{dE_p^3}$. In the limit $m\to 0$ (\ref{Asol}) reduces to
\begin{eqnarray}
A_1\to - \frac{1}{24\pi^2|\beta|^3}\,,\quad 
A_2\to -\frac{1}{8\pi^2|\beta|^3}\,,\quad 
A_3=0\,,
\label{Asolm0}
\end{eqnarray}
taking into account the first-order term \cite{Buzzegoli:2017cqy} Eq. (7.5) and (\ref{Asolm0}) we can write the formula for the axial current (\ref{jafull}) for case $m=0$ in the following form
\begin{eqnarray}
\langle j_{\mu}^{5}\rangle =\Big(\frac{1}{6}\big[T^2-\frac{\omega^2}{4\pi^2}\big]+\frac{\mu^2}{2\pi^2}-\frac{a^2}{8\pi^2}\Big)\omega_{\mu}+O(\varpi^5)\,.
\label{jasol}
\end{eqnarray}

Note again that according to \cite{Vilenkin:1980zv, Vilenkin:1979ui, Prokhorov:2017atp, Prokhorov:2018qhq} the third order in (\ref{jasol}) can be the last nonzero term. Since $A_3=0$, then, using the formulas for differentiation from \cite{Buzzegoli:2017cqy, Prokhorov:2017atp}, we get for (\ref{jasol})
\begin{eqnarray}
\partial^{\mu}\langle j_{\mu}^{5}\rangle =0\,.
\label{jadiv}
\end{eqnarray}

Thus, the axial charge in this approach is conserved in the massless limit, in contrast to \cite{Prokhorov:2017atp}.

\section{The density operator vs Wigner function}
\label{Sec:comparison}

In \cite{Prokhorov:2018qhq, Prokhorov:2017atp}, based on the Wigner function \cite{Becattini:2013fla}, the following general formula for the axial current in a nonstationary medium of massive fermions was obtained
\begin{eqnarray}
&& \langle j_{\mu}^{5}\rangle =
\frac{\omega_{\mu}+i\, \mathrm{sgn}(\omega a) a_{\mu}}{2(g_{\omega}-i g_a)}
\int\frac{d^3 p}{(2\pi)^3}\Big\{
n_{F}(E_p-\mu - g_{\omega}/2 + i g_a/2)-\nonumber \\
&& n_{F}(E_p-\mu + g_{\omega}/2 - i g_a/2)  
+n_{F}(E_p+\mu - g_{\omega}/2 + i g_a/2)-\nonumber \\
&& n_{F}(E_p+\mu + g_{\omega}/2 - i g_a/2)
\Big\}+ c.c.\,,
\label{jaWigner}
\end{eqnarray}
where
\begin{eqnarray} 
&& g_{\omega}=\frac{1}{\sqrt{2}}\big(\sqrt{(a^2-\omega^2)^2+4(\omega a)^2}+a^2-
\omega^2 \big)^{1/2}\,, \nonumber \\
&& g_a=\frac{1}{\sqrt{2}}\big(\sqrt{(a^2-\omega^2)^2+4(\omega a)^2}-a^2+\omega^2 \big)^{1/2}\,.
\label{gwga}
\end{eqnarray}

The formula (\ref{jaWigner}) was derived outside the perturbation theory.
In the limit $m=0$ for $T>\frac{g_a}{2\pi}$, (\ref{jaWigner}) leads to
\begin{eqnarray}
\langle j_{\mu}^{5}\rangle =\Big(\frac{1}{6}\big[T^2+\frac{a^{2} -
\omega^2}{4\pi^2}\big]+\frac{\mu^2}{2\pi^2}\Big)\omega_{\mu}+\frac{1}{12\pi^2}(\omega a)\,a_{\mu}\,.
\label{jaWignerm0}
\end{eqnarray}

For $a_{\mu}=0$ and passing to the comoving reference system, we obtain from (\ref{jaWigner})
\begin{eqnarray}
&& \langle  \bold{j}^{5}\rangle =
\int \frac{d^3 p}{(2\pi)^3}
\Big\{
n_{F}(E_p-\mu - \frac{\Omega}{2})-
n_{F}(E_p-\mu + \frac{\Omega}{2})+ \nonumber \\
&& n_{F}(E_p+\mu - \frac{\Omega}{2})-
n_{F}(E_p+\mu + \frac{\Omega}{2})
\Big\}\,\bold{e}_{\,\Omega}\,,
\label{jaWignera0}
\end{eqnarray}
where $\bold{e}_{\Omega}=\frac{\bold{\Omega}}{\Omega}$ is a unit vector along angular velocity. Let us first compare the formulas (\ref{jasol}) and (\ref{jaWignerm0}), which determine the axial current in the case of massless fermions. We see that the terms of the first order in $\omega$ coincide with each other and the standard formula for CVE, also the term $\omega^2\omega_{\mu}$ has the same coefficient, which also coincides with the result of \cite{Vilenkin:1979ui, Vilenkin:1980zv}. At the same time, the term $a^2\omega_{\mu}$ enters with different coefficients, and the term $(\omega a)\,a_{\mu}$ in (\ref{jasol}) is absent. Due to this, the axial charge is conserved for (\ref{jasol}) and is not conserved for (\ref{jaWignerm0}), where
\begin{eqnarray}
\partial^{\mu}\langle j_{\mu}^{5}\rangle = \partial^{\mu}\big[\frac{1}{12\pi^2}(\omega a)\,a_{\mu}\big]=\frac{1}{6\pi^2}(\omega a)(a^2 +\omega^2) \,.
\label{jadivWigner}
\end{eqnarray}

On the other hand, in formula (\ref{jaWignerm0}), unlike (\ref{jasol}), the combination of the form $\mu \pm (\Omega \pm ia)/2$ appears, since (\ref{jaWignerm0}) in the comoving frame of reference and for parallel angular velocity and acceleration $\bold{\Omega}||\bold{a}$ gives
\begin{eqnarray}
\langle\bold{j}^5\rangle=\Big(\frac{T^2\Omega}{6}+\frac{(\mu + \frac{\Omega}{2} + \frac{ia}{2})^3}{12\pi^2}-\frac{(\mu - \frac{\Omega}{2} - \frac{ia}{2})^3}{12\pi^2}+\frac{(\mu + \frac{\Omega}{2} - \frac{ia}{2})^3}{12\pi^2}-\frac{(\mu - \frac{\Omega}{2} +\frac{ia}{2})^3}{12\pi^2}\Big)\bold{e}_{\Omega}\,,
\label{jaWignerm0parallel}
\end{eqnarray}
which is a manifestation of the fact that the angular velocity and acceleration $a=| {\bf a}|$ play the role of chemical potentials, the latter being an imaginary one.

The possibility of appearance of such a combination in  (\ref{jaWignerm0}) can be seen already from the fact that 
$\mu^2$ and $\Omega^2$ enter with the same coefficient (after taking into account that $\omega^2=-\Omega^2 <0$) while the coeffcient of $| {\bf a}|^2= - a^2$ has the opposite sign. 

The fact that the imaginary chemical potential corresponds to acceleration leads, first, to the absence of terms of odd order in the acceleration in (\ref{jaWignerm0}), and also to the appearance of the Unruh temperature as the boundary temperature in the axial current, according to \cite{Prokhorov:2018qhq}.

Thus, both approaches give the same answer in the massless limit for the case of pure rotation $a_{\mu}=0$ and diverge when describing mixed effects (the terms $a^2\omega_{\mu}$ and $(\omega a)\,a_{\mu}$). 

In the more general case of massive fermions, the situation looks the same. In this case, it is necessary to compare the formulas (\ref{jaWignera0}) and (\ref{jafull}), (\ref{ja1}), (\ref{Asol}) (in advance it is clear that for $a_{\mu}\neq 0$, (\ref{jaWigner}) and (\ref{jafull}) are different). To do this, it is necessary to decompose (\ref{jaWignera0}) to the third order in $\Omega$. It is easy to show that the coefficients in this expansion are exactly given by the formula (\ref{Asol}) obtained by us for $A_1$ and (\ref{ja1}) (for (\ref{ja1})  the correspondence was shown in \cite{Buzzegoli:2017cqy}). Thus, (\ref{jaWignera0}) and (\ref{jafull}) coincide for $a_{\mu}=0$ in the first two non-vanishing orders in $\varpi$ and in the case of massive fermions. 

Note also that (\ref{jaWignerm0}) in the case of $T<\frac{g_a}{2\pi}$ contains additional corrections according to \cite{Prokhorov:2018qhq}, which are not included in (\ref{jasol}), since (\ref{jasol}) is obtained within the framework of perturbation theory. It should be expected that at temperatures below Unruh temperature, the behavior of the current either changes qualitatively, or the Unruh temperature sets the lower temperature boundary for accelerated moving systems according to \cite{Becattini:2017ljh, Prokhorov:2018qhq, Florkowski:2018myy}.

\section{Conclusions}
\label{Sec:conclusions}

Using the quantum statistical approach based on the equilibrium density operator (\ref{rho global}) we calculated the hydrodynamic coefficients in the axial current in the third order of perturbation theory in terms of the thermal vorticity tensor for the free Dirac fields.
Thus, we calculated the third-order corrections in the derivatives to the CVE.

The obtained expression coincides with the prediction based on the ansatz of the Wigner function in the first three orders of perturbation theory (formulas (\ref{jaWignera0}) and (\ref{jafull}), (\ref{ja1}), (\ref{Asol}) in the case of massive fermions and  (\ref{jaWignerm0}) and (\ref{jasol}) in the massless limit) for  $a_{\mu}=0$ and differ for $a_{\mu}\neq 0$. This indicates the correspondence of the two methods in describing the effects associated with pure rotation, and the discrepancy in describing mixed effects, when both rotation and acceleration are significant.

Effects in an axial current related to acceleration were investigated. In the approach with the Wigner function, as well as in the approach with the density operator, terms quadratic in acceleration appear. In the case of the Wigner function this is explained by the appearance of the combination $\mu \pm (\Omega \pm ia)/2$ - the appearance of an imaginary chemical potential associated with acceleration, forbids the appearance of odd acceleration terms. However, this combination does not arise in the approach with the density operator. The coefficients in front of the terms with acceleration in the two approaches differ. This leads, in particular, to the fact that the axial charge is conserved for the statistical operator and is not conserved for the Wigner function if there is an acceleration along the rotation axis in the system.

Note that the acceleration implies the non-equilibrium situation (c.f. \cite{Florkowski:2018ahw}) which might explain the discrepancy between Wigner function and density matrix approaches, both being the equilibrium ones. 
This problem, as well as other consequences of possible emerging instabilities and dissipation, require further investigation. 

{\bf Acknowledgments}

Useful discussions with V. Braguta, M. Buzzegoli, E. Grossi are gratefully acknowledged. The work was supported  by Russian Science Foundation Grant No 16-12-10059.


\appendix

\section{Calculation of quantum correlators}
\label{Sec:correlators}

Let's get formulas (\ref{Csol}), (\ref{Bsol}) for quantities $C^{\alpha_1\alpha_2|\alpha_3\alpha_4|\alpha_5\alpha_6|\lambda|ijk}$. Following \cite{Buzzegoli:2017cqy}, we represent all operators in (\ref{C}) in a split form. The operator $\mathcal{D}^{\alpha\beta}_{a b}(\partial_{X_1},\partial_{X_2})$, acting on the product of two Dirac fields, gives Belinfante energy-momentum tensor in the limit $X_1,X_2\to X$
\begin{eqnarray}
&&\hat{T}^{\alpha\beta}(X)=\lim_{\scriptscriptstyle X_1,X_2\to X} \mathcal{D}^{\alpha\beta}_{a b}(\partial_{X_1},\partial_{X_2})\bar{\Psi}_{a}(X_1)\Psi_{b}(X_2)\,,\nonumber \\ 
&&\mathcal{D}^{\alpha\beta}_{a b}(\partial_{X_1},\partial_{X_2})=\frac{i^{\delta_{0\alpha}+\delta_{0\beta}}}{4}
[\tilde{\gamma}_{a b}^{\alpha}(\partial_{X_2}-\partial_{X_1})^{\beta}+\tilde{\gamma}_{a b}^{\beta}(\partial_{X_2}-\partial_{X_1})^{\alpha}]\,,
\label{Tsplit}
\end{eqnarray}
and the axial current is expressed in terms of the operator $\mathcal{J}^{\lambda}_{5\, a b}$
\begin{eqnarray}
&&j^{\lambda}_5(X)=\lim_{\scriptscriptstyle X_1,X_2\to X} \mathcal{J}^{\lambda}_{5\, a b}\bar{\Psi}_{a}(X_1)\Psi_{b}(X_2)\,,\quad
\mathcal{J}^{\lambda}_{5\, a b}=i^{\delta_{0\lambda}-1}
(\tilde{\gamma}^{\lambda}\tilde{\gamma}^{5})_{a b}\,.
\label{jasplit}
\end{eqnarray}

Then taking into account (\ref{Tsplit}) and (\ref{jasplit}) we get
\begin{eqnarray}
&&\langle
T_{\tau}\hat{T}^{\alpha_1\alpha_2}(X)\hat{T}^{\alpha_3\alpha_4}(Y)\hat{T}^{\alpha_5\alpha_6}(Z)j_5^{\lambda}(0)\rangle_{\beta(x),c}=\lim_{
\def\arraystretch{0.5}\begin{array}{ll}
{\scriptscriptstyle X_1,X_2\to X}\vspace{0.1mm}\\
{\scriptscriptstyle Y_1,Y_2\to Y}\\
{\scriptscriptstyle Z_1,Z_2\to Z}\\
{\scriptscriptstyle F_1,F_2\to F=0}
\end{array}} \mathcal{D}^{\alpha_1\alpha_2}_{a_1 a_2}(\partial_{X_1},\partial_{X_2}) \nonumber \\
&&\mathcal{D}^{\alpha_3\alpha_4}_{a_3 a_4}(\partial_{Y_1},\partial_{Y_2})\mathcal{D}^{\alpha_5\alpha_6}_{a_5 a_6}(\partial_{Z_1},\partial_{Z_2})\mathcal{J}^{\lambda}_{5\, a_7 a_8}
\langle T_{\tau} \bar{\Psi}_{a_1}(X_1)\Psi_{a_2}(X_2)\bar{\Psi}_{a_3}(Y_1)\Psi_{a_4}(Y_2) \nonumber \\ 
&&\bar{\Psi}_{a_5}(Z_1)\Psi_{a_6}(Z_2)\bar{\Psi}_{a_7}(F_1)\Psi_{a_8}(F_2)\rangle_{\beta(x),c}\,.
\label{corrsplit}
\end{eqnarray}

Using Wick theorem, the calculation of averages in (\ref{corrsplit})  can be reduced to finding the means of the quadratic combinations of Dirac fields of the form $\langle T_{\tau}\Psi_{a_1}(X_1)\bar{\Psi}_{a_2}(X_2)\rangle_{\beta(x)}$, which are thermal propagators. Leaving only the connected correlators, we obtain
\begin{eqnarray}
&&\langle T_{\tau}\bar{\Psi}_{a_1}(X_1)\Psi_{a_2}(X_2)\bar{\Psi}_{a_3}(Y_1)\Psi_{a_4}(Y_2)\bar{\Psi}_{a_5}(Z_1)\Psi_{a_6}(Z_2)\bar{\Psi}_{a_7}(F_1)\Psi_{a_8}(F_2)\rangle_{\beta(x),c}=\nonumber \\ 
&&-\bar{G}_{a_1a_4}(X_1,Y_2)G_{a_2a_5}(X_2,Z_1)\bar{G}_{a_3a_8}(Y_1,F_2)G_{a_6a_7}(Z_2,F_1)+\nonumber \\ 
&&\bar{G}_{a_1a_4}(X_1,Y_2)G_{a_2a_7}(X_2,F_1)\bar{G}_{a_3a_6}(Y_1,Z_2)\bar{G}_{a_5a_8}(Z_1,F_2)-\nonumber \\ 
&&\bar{G}_{a_1a_6}(X_1,Z_2)G_{a_2a_3}(X_2,Y_1)G_{a_4a_7}(Y_2,F_1)\bar{G}_{a_5a_8}(Z_1,F_2)-\nonumber \\ 
&&\bar{G}_{a_1a_6}(X_1,Z_2)G_{a_2a_7}(X_2,F_1)\bar{G}_{a_3a_8}(Y_1,F_2)G_{a_4a_5}(Y_2,Z_1)+\nonumber \\ 
&&\bar{G}_{a_1a_8}(X_1,F_2)G_{a_2a_3}(X_2,Y_1)G_{a_4a_5}(Y_2,Z_1)G_{a_6a_7}(Z_2,F_1)-\nonumber \\ 
&&\bar{G}_{a_1a_8}(X_1,F_2)G_{a_2a_5}(X_2,Z_1)\bar{G}_{a_3a_6}(Y_1,Z_2)G_{a_4a_7}(Y_2,F_1)\,,
\label{vik}
\end{eqnarray}
where the thermal propagators $G_{a_1a_2}(X_1,X_2)=\langle T_{\tau}\Psi_{a_1}(X_1)\bar{\Psi}_{a_2}(X_2)\rangle_{\beta(x)}$, and $\bar{G}_{a_1a_2}(X_1,X_2)=\langle T_{\tau}\bar{\Psi}_{a_1}(X_1)\Psi_{a_2}(X_2)\rangle_{\beta(x)}$ have the standard form \cite{Buzzegoli:2017cqy, Laine:2016hma, Kapusta:2006pm}
\begin{eqnarray}
G_{a_1a_2}(X_1,X_2)=\SumInt_{\{P\}}e^{iP^+(X_1-X_2)}(-i\slashed{P}^{+}+m)_{a_1a_2}\Delta(P^+)\,,
\label{prop}
\end{eqnarray}
and, respectively, for $\bar{G}$. In (\ref{prop}) we introduce the notation
\begin{eqnarray}P^\pm=(p^{\pm}_n,\bold{p})\,,\,
p^{\pm}_n=\pi(2n+1)/|\beta|\pm\mu\,,\,
\SumInt_{\{P\}}=\frac{1}{|\beta|}\sum_{n=-\infty}^{\infty}\int\frac{d^3p}{(2\pi)^3}\,,\,
\Delta(P)=\frac{1}{P^2+m^2}\,.
\label{not}
\end{eqnarray}

In $\Delta(P)$  the square is taken with the Euclidean metrics, as in $\slashed{P}^{+}=P_{\mu}^+\tilde{\gamma}_{\mu}$ (unlike from $P^+(X_1-X_2)$, where the metrics is non-Euclidean in accordance with \cite{Laine:2016hma}).

Now substitute (\ref{prop}) in (\ref{vik}) and then in (\ref{corrsplit}). Then we differentiate in operators $\mathcal{D}^{\alpha\beta}_{a b}(\partial_{X_1},\partial_{X_2})$, group the matrices in the form of a trace, taking into account that in the exponential factor we can cancel terms with chemical potential. For simplicity, let us analyze the transformations for the first term in (\ref{vik})
\begin{eqnarray}
&&-\lim_{
\def\arraystretch{0.5}\begin{array}{ll}
{\scriptscriptstyle X_1,X_2\to X}\vspace{0.1mm}\\
{\scriptscriptstyle Y_1,Y_2\to Y}\\
{\scriptscriptstyle Z_1,Z_2\to Z}\\
{\scriptscriptstyle F_1,F_2\to F=0}
\end{array}} \mathcal{D}^{\alpha_1\alpha_2}_{a_1 a_2}(\partial_{X_1},\partial_{X_2}) \mathcal{D}^{\alpha_3\alpha_4}_{a_3 a_4}(\partial_{Y_1},\partial_{Y_2})\mathcal{D}^{\alpha_5\alpha_6}_{a_5 a_6}(\partial_{Z_1},\partial_{Z_2})\mathcal{J}^{\lambda}_{5\, a_7 a_8}\bar{G}_{a_1a_4}(X_1,Y_2)\nonumber \\
&&G_{a_2a_5}(X_2,Z_1)\bar{G}_{a_3a_8}(Y_1,F_2)G_{a_6a_7}(Z_2,F_1)=
-\SumInt_{\{P,Q,K,R\}}e^{-i\bold{p}(\bold{x}-\bold{y})-i\bold{q}(\bold{x}-\bold{z})-i\bold{k}\bold{y}-i\bold{r}\bold{z}}\nonumber \\
&&e^{ip^{-}_n(\tau_x -\tau_y)+iq^{+}_n(\tau_x -\tau_z)+ik^{-}_n \tau_y +ir^{+}_n\tau_z}\Delta(P^{-})\Delta(Q^{+})\Delta(K^{-})\Delta(R^{+})\nonumber \\
&&\mathrm{tr}\Big[
(i\slashed{K}^{-}+m)\mathcal{D}^{\alpha_3\alpha_4}(iK^{-},-iP^{-})
(i\slashed{P}^{-}+m)\mathcal{D}^{\alpha_1\alpha_2}(iP^{-},iQ^{+})
(-i\slashed{Q}^{+}+m)\nonumber \\
&&\mathcal{D}^{\alpha_5\alpha_6}(-iQ^{+},iR^{+})
(-i\slashed{R}^{+}+m)\mathcal{J}^{\lambda}_{5}
\Big]\,.
\label{C1a}
\end{eqnarray}

Summing over the Matsubara frequencies in (\ref{C1a}) using the formula \cite{Buzzegoli:2017cqy, Laine:2016hma}
\begin{eqnarray}
\frac{1}{|\beta|}\sum_{\omega_n}\frac{(\omega_n\pm i\mu)^k e^{i(\omega_n\pm i\mu)\tau}}{(\omega_n\pm i\mu)^2+E^2}=\frac{1}{2E}\sum_{s=\pm 1}(-isE)^k e^{\tau s E}[\theta(-s\tau)-n_F(E\pm s\mu)]\,,
\label{sum}
\end{eqnarray}
we obtain
\begin{eqnarray}\label{C1b}
&&-\frac{1}{16}\int\sum_{\def\arraystretch{0.5}\begin{array}{ll}
{\scriptscriptstyle s_1,s_2,s_3,}\\
{\scriptscriptstyle s_4=\pm 1}
\end{array}} \frac{d^3pd^3qd^3kd^3r}{(2\pi)^{12}E_{p}E_{q}E_{k}E_{r}}
e^{(\tau_x-\tau_y)s_1 E_{p}+(\tau_x-\tau_z)s_2 E_{q}+\tau_y s_3 E_{k}+\tau_z s_4 E_{r}}
\\ \nonumber 
&&e^{-i\bold{p}(\bold{x}-\bold{y})-i\bold{q}(\bold{x}-\bold{z})-i\bold{k}\bold{y}-i\bold{r}\bold{z}}
B_{-+-+,(\tau_x-\tau_y),(\tau_x-\tau_z)}^{\alpha_3\alpha_4\alpha_1\alpha_2\alpha_5\alpha_6\lambda}(\widetilde{K},\widetilde{K},-\widetilde{P},\widetilde{P},\widetilde{P},\widetilde{Q},-\widetilde{Q},-\widetilde{Q},\widetilde{R},-\widetilde{R})\,,
\end{eqnarray}
where the quantities $B$ are given by (\ref{Bsol}). Substituting (\ref{C1b}) in (\ref{C}) and using formula
\begin{eqnarray}\nonumber
&&\int d^3p d^3q d^3k d^3r d^3x d^3y d^3z \,f(\bold{p},\bold{q},\bold{k},\bold{r})e^{-i\bold{p}(\bold{x}-\bold{y})
-i\bold{q}(\bold{x}-\bold{z})-i\bold{k}\bold{y}-i\bold{r}\bold{z}} x^{i}y^{j}z^{k}=\\
&&i(2\pi)^9\int d^3p \Big(\frac{\partial^3}{\partial r^{k}\partial k^{j}\partial p^{i}}+\frac{\partial^3}{\partial r^{k}\partial k^{j}\partial k^{i}}\Big)f(\bold{p},\bold{q},\bold{k},\bold{r})\Big|\def\arraystretch{0.5}\begin{array}{ll}
{\scriptstyle \bold{q}=-\bold{p}}\\
{\scriptstyle \bold{k}=\bold{p}}\\
{\scriptstyle \bold{r}=-\bold{p}}\,,
\end{array}
\label{difeq}
\end{eqnarray}
following from the properties of the delta function, we finally obtain
\begin{eqnarray} 
&&-\frac{i}{128\pi^3|\beta|^3}\int\sum_{\def\arraystretch{0.5}\begin{array}{ll}
{\scriptscriptstyle s_1,s_2,s_3,}\vspace{0.1mm}\\
{\scriptscriptstyle s_4=\pm 1}
\end{array}} d\tau_x d\tau_y d\tau_z p^2 dp \sin(\theta) d\theta d\varphi \nonumber \\ 
&&\Big(\frac{\partial^3}{\partial r^{k}\partial k^{j}\partial p^{i}}+\frac{\partial^3}{\partial r^{k}\partial k^{j}\partial k^{i}}\Big)
\frac{1}{E_{p}E_{q}E_{k}E_{r}}
e^{(\tau_x-\tau_y)s_1 E_{p}+(\tau_x-\tau_z)s_2 E_{q}+\tau_y s_3 E_{k}+
\tau_z s_4 E_{r}} \nonumber \\
&&B_{-+-+,(\tau_x-\tau_y),(\tau_x-\tau_z)}^{\alpha_3\alpha_4\alpha_1\alpha_2\alpha_5\alpha_6\lambda}(\widetilde{K},\widetilde{K},-\widetilde{P},\widetilde{P},\widetilde{P},\widetilde{Q},-\widetilde{Q},-\widetilde{Q},\widetilde{R},-\widetilde{R})\Big|\def\arraystretch{0.5}
\begin{array}{ll}
{\scriptstyle \bold{q}=-\bold{p}}\\
{\scriptstyle \bold{k}=\bold{p}}\\
{\scriptstyle \bold{r}=-\bold{p}}
\end{array}\,,
\label{C1c}
\end{eqnarray}
which corresponds to the first term in (\ref{Csol}). Performing transformations from (\ref{C1a}) to (\ref{C1c}) with other terms in (\ref{vik}), we obtain (\ref{Csol}).

\end{document}